\documentclass[runningheads]{llncs}
\usepackage{graphicx}
\graphicspath{ {./img/} }
\begin{document}
\title{Deep learning for spike detection in deep brain stimulation surgery}
%
%
\author{Arkadiusz Nowacki\inst{1} \and
Ewelina Kołpa\inst{1} \and
Mateusz Szychiewicz\inst{1} ,\\
Konrad Ciecierski\inst{2}}
\authorrunning{A. Nowacki et al.}
%
\institute{Department of Electronics and Information Technology, Warsaw University of Technology\\
\email{arkadiusz.nowacki.stud@pw.edu.pl, ewelina.kolpa.stud@pw.edu.pl, mateusz.szychiewicz.stud@pw.edu.pl}
\and
Research and Academic Computer Network (NASK-PIB)
\email{konrad.ciecierski@nask.pl}}
\maketitle              
\begin{abstract}
Deep brain stimulation (DBS) is a neurosurgical procedure successfully used to treat conditions such as Parkinson's disease. Electrostimulation, carried out by implanting electrodes into an identified focus in the brain, makes it possible to reduce the symptoms of the disease significantly. In this paper, a method for analyzing recordings of neuronal activity acquired during DBS neurosurgery using deep learning is presented. We tested using a convolutional neural network (CNN) for this purpose. Based on the time window, the classifier assesses whether neuronal activity (spike) is present. The maximum accuracy value for the classifier was 98.98\%, and the area under the receiver operating characteristic curve (AUC) was 0.9898. The method made it possible to obtain a classification without using data preprocessing.

\end{abstract}
\section{Introduction and Problem Formulation}

Deep brain stimulation (DBS) is an efficient method in the field of neurosurgery, which not only can be used to treat Parkinson's disease but also Tourette's syndrome, movement, and anxiety disorders \cite{casagrande2019brain}. This is a more efficient method for the localization of a small structure Subthalamic Nucleus than standard imaging techniques such as CT and MRI. The main essence of DBS is the modulation of specific structures in the brain by electrical impulses generated with a frequency of 100-200 Hz via surgically implanted electrodes in specific brain regions. These electrodes emit electrical impulses that can modulate the activity of neurons, which can help to alleviate symptoms of conditions such as dystonia and essential tremor. Due to its clinical effectiveness, scientists are still looking for alternative ways to use it in psychiatry and other fields of medicine. 
To ensure that the electrodes are placed in the optimal location, doctors use spike detection to identify the specific neural activity that is associated with the condition being treated.

Once the electrodes are in place, spike detection can also be used to monitor the effects of DBS treatment over time. By measuring the neural activity before and after DBS, doctors can determine if the treatment has the desired effect and adjust the stimulation parameters as needed. Additionally, spike detection can be used to detect any side effects of DBS treatment, such as changes in cognitive function or mood.

The following parts of this work will discuss related scientific studies on data acquisition, cleaning of artifacts, and spike detection.
The developed methods of spike detection using deep learning, data analysis on which experiments were carried out, and their results will also be described. The paper proposes the use of a convolutional neural network for spike detection. The model was trained on real data obtained during the DBS operation. Additionally, the impact of the training data on accuracy, precision, recall, and F1 score is considered.

\section{Related Work}

\subsection{Spike detection}
A neural spike, also known as an action potential, is a brief, rapid change in the electrical potential of a neuron. This is due to the rapid influx of positively charged ions into the cell, causing the membrane potential to change rapidly from negative to positive\cite{quian2009extracting}.

Spike detection identifies the presence of nerve impulses in an electrophysiological signal, such as EEG or extracellular recording. There are various methods for detecting spikes in neural signals, including:

\begin{enumerate}
  \item threshold-based methods
  \item wavelet-based methods
  \item template matching
  \item spectral method
\end{enumerate}
Threshold-based methods involve setting a threshold value; any signal exceeding that threshold is considered a spike.

Wavelet-based methods use wavelet transforms to decompose the signal into different frequency bands and then identify spikes based on wavelet coefficients.

Template matching involves creating a peak waveform template and then using that template to detect signal spikes.

The spectral method is based on the power spectral density of signals and is used to detect high-frequency spikes\cite{wilson2002spike}.

\subsection{Deep learning in brain waves analysis}
Deep learning is a subfield of machine learning that uses neural networks with multiple layers to analyze complex data. Much work has been done on EEG analysis in terms of using deep learning to analyze brainwaves.
In the context of brainwave analysis, deep learning techniques such as convolutional neural networks (CNNs) and recurrent neural networks (RNNs) can be used to analyze large amounts of electroencephalography data, identifying patterns and features that are indicative of different cognitive states\cite{schirrmeister2017deep}.

Convolutional neural networks (CNNs) are particularly well suited for analyzing electroencephalography data, as they can learn spatial representations of the data by applying filters to small regions of the input data. This allows them to identify patterns in the electroencephalography data specific to certain brain regions or cognitive states\cite{schirrmeister2017deep}.

Recurrent neural networks (RNNs) are another type of deep learning algorithm that can be used in brainwave analysis. RNNs can process sequential data, such as time series data, and learn patterns in the data that span multiple time steps. This makes them well suited for analyzing electroencephalography data, as signals are time series data that change over time\cite{alhagry2017emotion}.

Deep learning techniques such as deep belief networks (DBNs) and autoencoders (AEs) are also used for EEG-based brain-computer interfaces (BCIs) and for identifying abnormal brain activities such as seizures and sleep disorders\cite{roy2019deep}.

\section{Deep Learning-based Algorithm for Spike Detection}

\subsection{Input Dataset Description}
The analyzed data came from recordings made during deep brain stimulation surgery. The sampling device took data at a frequency of 24 kHz. Such frequency means that there are 24 samples per 1 ms of recording. Each recording is 4s long, giving 240000 samples per recording. To create the dataset, the data was processed to obtain the timestamps in which the spikes occurred.

The recordings contain a lot of noise and interference. To better detect spikes, has been carried out data renormalization. The renormalization was based on the median absolute deviation (MAD). MAD is calculated by finding the median of a data set and then finding the absolute difference between each data point and the median. The median of absolute differences is then taken as a measure of the dispersion or variability of the data set\cite{hoaglin1983understanding}.

\begin{figure}[h]
    \centering
    \includegraphics[width=0.45\textwidth]{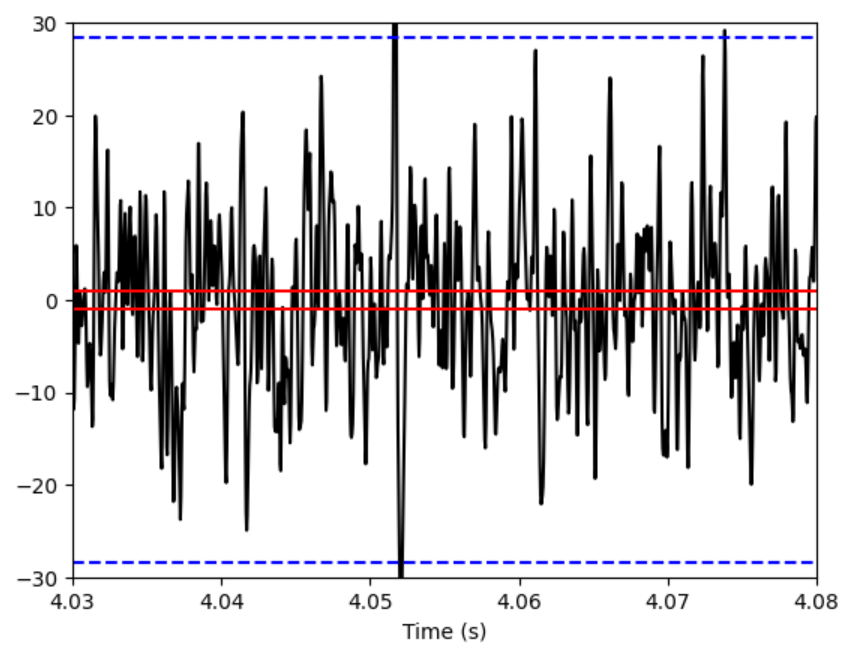}
    \includegraphics[width=0.45\textwidth]{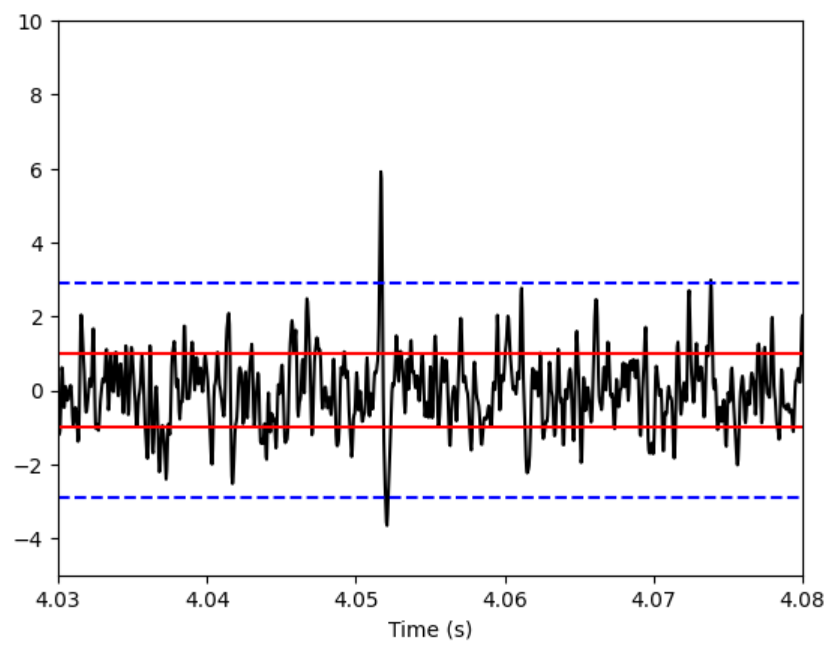}
    \caption{Samples from the recording before (left) and after (right) renormalization. In red: +/- the MAD; in dashed blue +/- the SD}
    \label{fig:record_sample}
\end{figure}

\begin{equation}
{MAD} = {median} (|X-{median}(X)|)\frac{1}{0.6745}
\end{equation}

Where:
\begin{itemize}
    \item $X$: is the dataset
    \item $median(X)$: is the median of the dataset
    \item $|x - median(x)|$: is the absolute difference between each data point and the median of the dataset
\end{itemize}

The constant $\frac{1}{0.6745}$ is used to make the MAD comparable to the standard deviation for a normal distribution\cite{rousseeuw1993alternatives}. 

Renormalization aims to rescale the raw data such that the standard deviation (SD) noise is approximately 1. Exact scaling may not be feasible, but MAD significantly approximates the noise value to SD (Figure \ref{fig:record_sample}). Renormalization using MAD can help when recordings are multichannel. This allows for comparing electrode/channel values with each other.

\begin{figure}[h]
    \centering
    \includegraphics[width=0.65\textwidth]{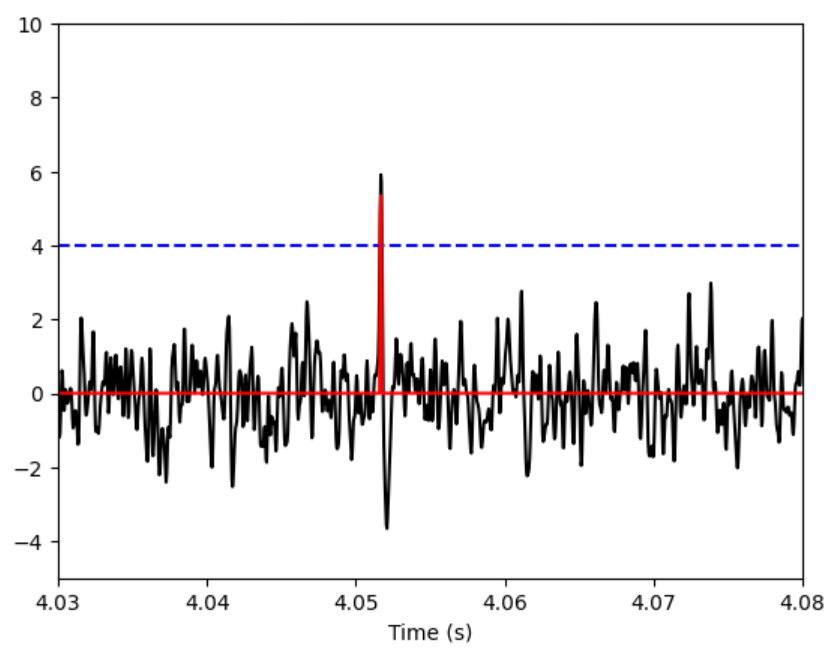}
    \caption{Sample from the recording with detection threshold (dashed blue) and the filtered and rectified trace (red).}
    \label{fig:preprocessed_sample}
\end{figure}

Spike detection involves selecting local extremes above a designated threshold. The data is first filtered using a box filter (a moving average) to reduce high-frequency noise\cite{quiroga2004unsupervised}. The box filter works by averaging the data over a specific time window, smoothing out high-frequency signal fluctuations. This can help improve spike detection because it reduces the impact of noise on the detection process.

The SD value determines the detection threshold. It can be a multiple of SD or its value. Based on the crossing of this threshold, sites where a spike could occur, are selected(Figure \ref{fig:preprocessed_sample}). 

\begin{figure}[h]
    \centering
    \includegraphics[width=0.65\textwidth]{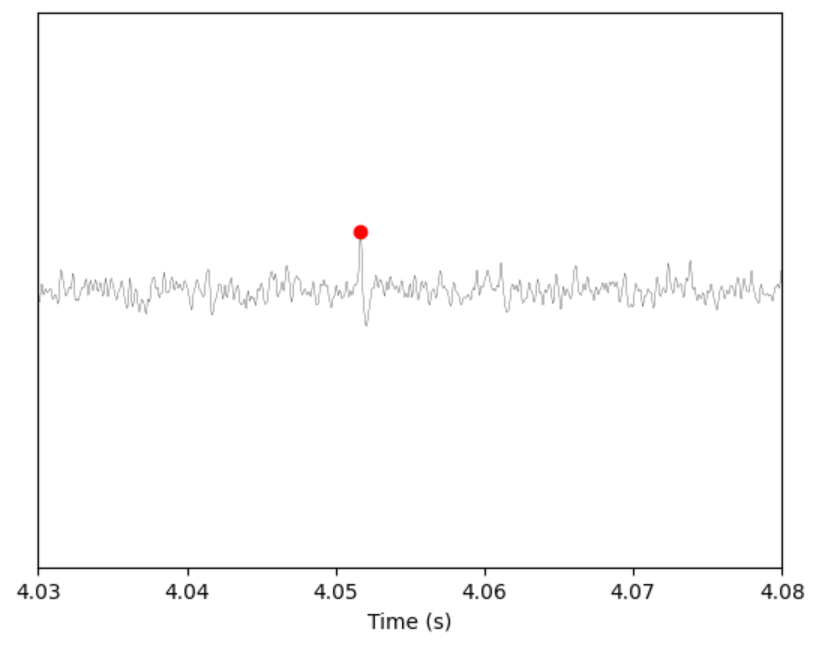}
    \caption{Raw data with a detected spike (red dot)}
    \label{fig:detected_spike}
\end{figure}

Spike detection involves checking selected samples. The absolute value of detected spikes and their minimum distances from each other is checked\cite{quiroga2004unsupervised}. This makes it possible to filter out distorted, overlapping spikes. This way, timestamps were obtained where spikes occur (Figure \ref{fig:detected_spike}).

\begin{figure}[h]
    \centering
    \includegraphics[width=0.65\textwidth]{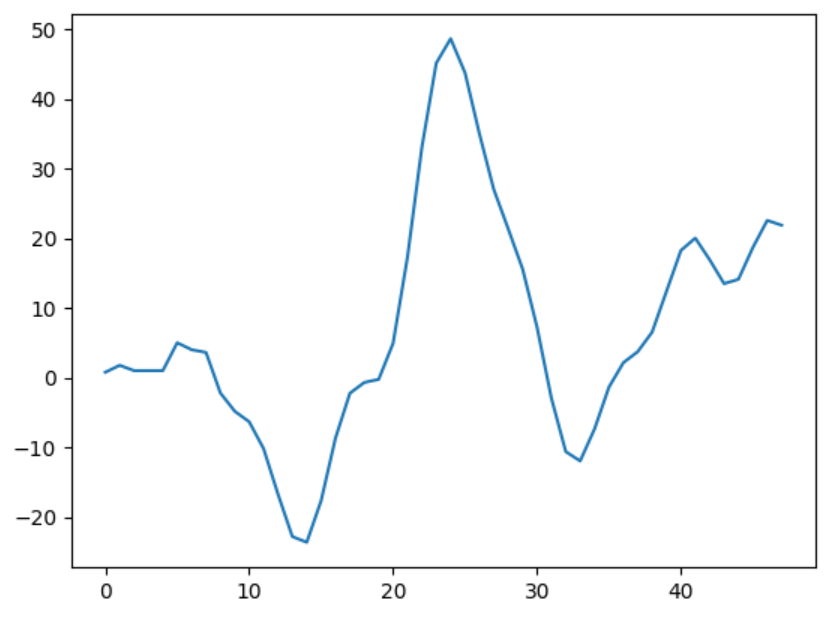}
    \caption{Sample time window with spike}
    \label{fig:time_window_spike}
\end{figure}

Spikes usually last about one millisecond\cite{kandel2000principles}. To create a training and validation dataset, time windows were stretched around the timestamps. As the device operates at a frequency of 24 samples per millisecond, time windows of 48 samples were created.
The time windows were created using the raw data before renormalization (Figure \ref{fig:time_window_spike}). The resulting data was divided into a training and validation dataset at a ratio of 80:20.

\subsection{Neural network}

\begin{figure}[h]
    \centering
    \includegraphics[width=1\textwidth]{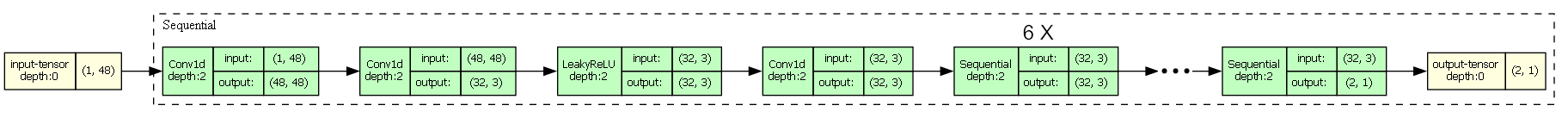}
    \includegraphics[width=1\textwidth]{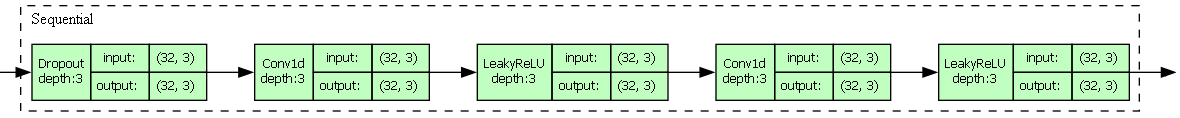}
    \caption{Structure of the neural network (top) and the structure of the sequence, repeated six times (bottom)}
    \label{fig:neural_network}
\end{figure}

In order to detect spikes, was created a binary classifier. It was based on a convolutional neural network. Using deep learning with convolutional neural networks yielded promising results in classifying EEG problems\cite{schirrmeister2017deep}. The architecture of the created network is shown in Figure \ref{fig:neural_network}.

The size of the input data corresponds to the size of the time window. The recordings are single-channel, and the window is 48 samples in size, so the input vector is 1x48. The input data is then preprocessed through several one-dimensional convolution layers. The Leaky Rectified Linear Unit (or Leaky ReLU) is used here. Unlike the standard ReLU, the function has a slight bias for negative values, unlike the ReLU, where for negative values, the function value is equal to 0\cite{xu2020reluplex}.

A block was then applied: dropout, a one-dimensional convolutional layer, and Leaky ReLU. The block is designed to get all the necessary information regarding the time window and the presence of a spike in it. This block is repeated six times.

The data is then modified to the target output size. The output is of form 2x1. The two outputs correspond to binary classification: no spike(or presence of noise) and the presence of spike. To process the data in this way, two layers were used: a one-dimensional convolutional network and a sigmoid. Sigmoid was used to normalize the data to an interval of 0-1 so that the largest value\cite{han1995influence}. The largest value thus indicates the classifier's prediction.

\section{Numerical Experiments and Performance Evaluation}

\subsection{Neural Network Training}

The classifier was trained for 15 epochs. For every epoch, a checkpoint of the trained model was created, and its metric values were saved. The loss function is Binary Cross Entropy, and the learning rate was set to 0.001; adaptive Moment Estimation (Adam) was chosen as an optimizer.

The following metrics have been checked:
\begin{enumerate}
  \item accuracy
  \item precision
  \item recall
  \item $F_{1}$ score
\end{enumerate}

Accuracy is a commonly used metric to evaluate the performance of a classification model. It is the ratio of the number of correct predictions made by the model to the total number of predictions made. Accuracy is between 0 and 1, where 1 is perfect accuracy, and 0 is no accuracy\cite{alpaydin2020introduction}.

\begin{equation}
{ACC} = \frac{TP + TN}{TP + TN + FN + FP}
\end{equation}

Precision is the ratio of the number of true positive predictions (i.e., the number of times the model correctly predicted a positive class) to the total number of positive predictions made by the model. Precision is between 0 and 1, where 1 mean perfect precision and 0 mean no precision\cite{alpaydin2020introduction}.

\begin{equation}
{Precision} = \frac{TP}{TP + FP}
\end{equation}

The recall is the ratio of the number of true positive predictions (i.e., the number of times the model correctly predicted the positive class) to the total number of positive instances in the dataset. The recall is between 0 and 1, where 1 represents perfect recall, and 0 represents no recall\cite{alpaydin2020introduction}.

\begin{equation}
{Recall} = \frac{TP}{TP + FN}
\end{equation}

The $F_{1}$ score balances precision and recall, giving equal weight to both. A high F1 score means the model has both a high precision and recall.

\begin{equation}
{F_{1}} = 2 * \frac{Precision * Recall}{Precision + Recall}
\end{equation}

Before the actual training, the number of blocks containing dropout, a one-dimensional convolutional layer, and Leaky ReLU was selected. The average values of the accuracy metrics were compared for 15 epochs. Values have been compared in the tabular \ref{table:blocks_compare}. 

\begin{table}[h]
    \caption{Comparison of average accuracy metrics against a number of blocks.}
    \centering
    \begin{tabular}{cc}
        \hline
        Number block & Accuracy (avg)\\
        \hline
        1 & 0.9686\\
        3 & 0.9726 \\
        6 & 0.9827 \\
        9 & 0.9829 \\
        12 & 0.9831 \\
        \hline
    \end{tabular}
    \label{table:blocks_compare}
\end{table}
Based on the comparison, six blocks were selected. Admittedly, nine and 12 achieved higher average accuracy but significantly increased training and prediction times, with minimal accuracy gains. 

The training was done using a previously created dataset discussed in the previous chapter.

\subsection{Results of Experiments}
Classifier training was carried out in several variants, using 25\%, 50\%, 75\%, and 100\% of the training set. It was thus tested how much data the network needs to achieve promising results. The results for the best models are presented in table \ref{table:percent_compare}.

\begin{table*}[ht]
    \caption{Compare the metrics against the percentage of the original training set used (the best results)}
    \centering
    \begin{tabular}{p{0.20\linewidth}p{0.20\linewidth}p{0.20\linewidth}p{0.20\linewidth}p{0.20\linewidth}}
    \hline
    Metric & 25\% data & 50\% data & 75\% data & 100\% data\\
    \hline
    Accuracy & 0.9099 & 0.9852 & 0.9865 & \textbf{0.9898}\\
    Precision & \textbf{0.9963} & 0.9929 & 0.9740 & 0.9846\\
    Recall & 0.8226 & 0.9774 & \textbf{0.9996}   & 0.9951\\
    $F_{1}$ score & 0.9011 & 0.9851 & 0.9866  & \textbf{0.9898}\\
    \hline
    \end{tabular}
    \label{table:percent_compare}
\end{table*}

The best results were achieved by the model for which the entire available training set was used. The value of the accuracy metric was 0.9898, and the F1 score was 0.9898 (both values were the highest in the entire comparison).

It is worth noting that high metrics values were also achieved by models trained at 50\% and 75\% of the original training set.

\begin{figure}[h]
    \centering
    \includegraphics[width=0.45\textwidth]{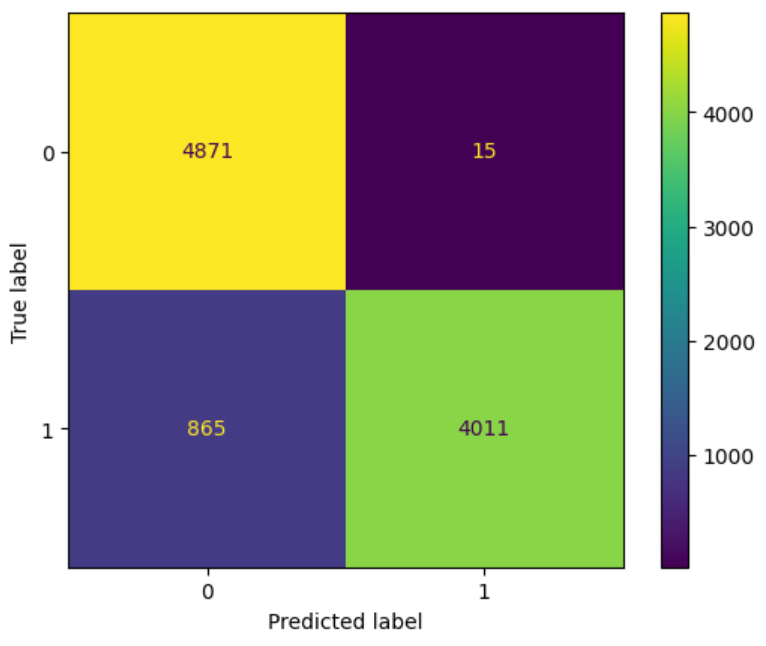}
    \includegraphics[width=0.45\textwidth]{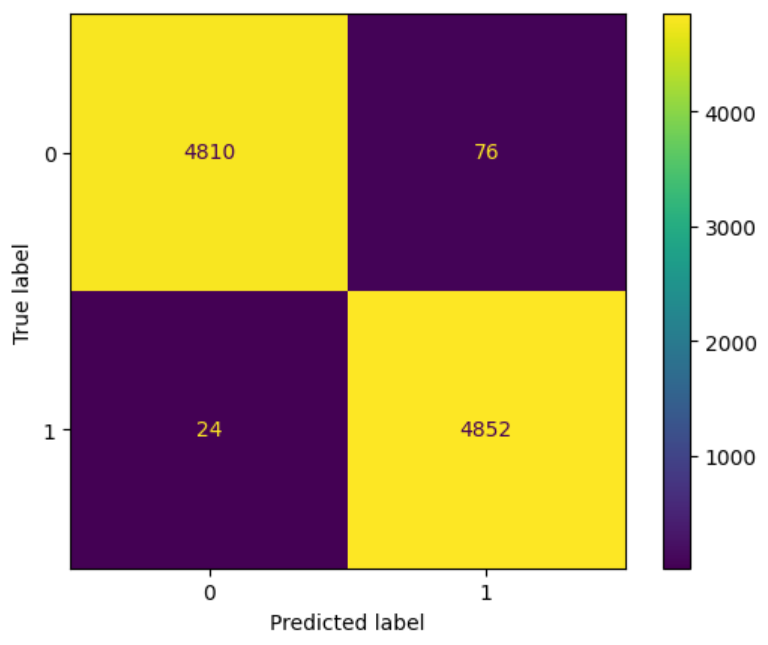}
    \caption{Confusion matrixes for 25\% (left) and 100\% (right) dataset (the best model)}
    \label{fig:confusion_matrix}
\end{figure}

Confusion matrices were created to illustrate the results (Figure \ref{fig:confusion_matrix}). The validation set consisted of 9762 examples. In the case of the best model, it achieved high values for True Positive (4852) and True Negative (4810). This gives the correct answer for 98\% and over 99\% for true negative and true positive, respectively.

\begin{figure}[h]
    \centering
    \includegraphics[width=0.65\textwidth]{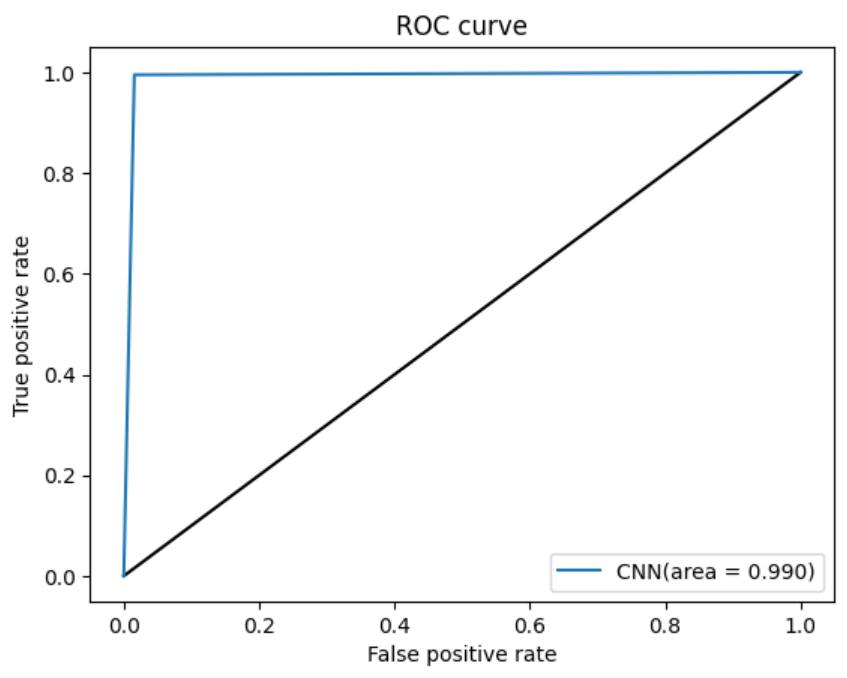}
    \caption{ROC curve for best model}
    \label{fig:roc}
\end{figure}

In order to check the correctness of distinguishing between classes by the classifier, a graph of the ROC curve was created (Figure \ref{fig:roc}). The area under the curve (AUC) was 0.990.

\section{Summary and Conclusions}
In this paper, we presented the use of deep learning using convolutional neural networks to detect spikes in recordings from DBS surgery automatically. The best model achieved 0.9898 accuracies and $F_{1}$ score of 0.9898. The binary classifier has yielded promising results for use in accelerating automatic spike detection. Reducing the training crop by 50\% yielded further satisfactory results.

A high value of the AUC metric (0.990) means that the classifier is good at recognizing individual classes. It can distinguish spikes from noise contained in the recording or overlapping spikes.

The classifier can be used to find spikes for sorting faster, and can also search for correct spikes in a distorted narrative containing much noise.

%
%
%
%
\bibliographystyle{splncs04}
\bibliography{references}
\end{document}